\documentclass[superscriptaddress,twocolumn,showpacs,preprintnumbers,amsmath,amssymb]{revtex4} 
\usepackage[dvipdfm]{graphicx}% Include figure files
\usepackage{dcolumn,upgreek}% Align table columns on decimal point
\usepackage{bm}% bold math
\usepackage{color}
\def\ii{{\mathrm{i}}}
\def\ee{{\mathrm{e}}}
\def\dd{{\mathrm{d}}}

\def\no{{\nonumber}} %\nonumber

\def\ket#1{|#1\rangle}

\def\bracketii#1#2#3{\langle #1 | #2| #3\rangle}

\begin{document}

%\preprint{APS/123-QED}

\title{Time-reversed two-photon interferometry for phase super-resolution}% Force line breaks with \\

\author{Kazuhisa Ogawa}
\email{ogawa@giga.kuee.kyoto-u.ac.jp}
\affiliation{% 
Department of Electronic Science and Engineering, Kyoto University, Kyoto 615-8510, Japan
}%
 
\author{Shuhei Tamate}
\affiliation{% 
Center for Emergent Matter Science, RIKEN, Wako-shi, Saitama 351-0198, Japan
}%
\author{Hirokazu Kobayashi}
\affiliation{%
Department of Electronic and Photonic System Engineering,\\
Kochi University of Technology, Tosayamada-cho, Kochi 782-8502, Japan
}%
\author{Toshihiro Nakanishi}
\affiliation{% 
Department of Electronic Science and Engineering, Kyoto University, Kyoto 615-8510, Japan
}%
\author{Masao Kitano}
\affiliation{% 
Department of Electronic Science and Engineering, Kyoto University, Kyoto 615-8510, Japan
}%

\date{\today}% It is always \today, today,

\begin{abstract}

We observed two-photon phase super-resolution in an unbalanced Michelson interferometer with classical Gaussian laser pulses.
Our work is a time-reversed version of a two-photon interference experiment using an unbalanced Michelson interferometer. 
A measured interferogram exhibits two-photon phase super-resolution with a high visibility of $97.9\%\pm0.4\%$.
Its coherence length is about 22 times longer than that of the input laser pulses.
It is a classical analogue to the large difference between the one- and two-photon coherence lengths of entangled photon pairs.

\end{abstract}

\maketitle

\section{Introduction}\label{sec:5_0}

Entangled photon pairs generated by spontaneous parametric down-conversion (SPDC) have peculiar characteristics that have never been seen in classical optics.
The time-frequency correlation is one such characteristic.
Two time-frequency correlated photons tend to be detected simultaneously, and the sum of their frequencies is constant. 
The two-photon coherence length of time-frequency entangled photon pairs is much larger than that of individual photons. 
By utilizing these properties, various two-photon interference phenomena have been observed, such as automatic dispersion cancellation \cite{PhysRevA.45.6659,PhysRevLett.68.2421}, nonlocal interference \cite{PhysRevLett.62.2205,PhysRevLett.65.321}, and two-photon phase super-resolution in an unbalanced Michelson interferometer \cite{PhysRevLett.66.1142,PhysRevLett.86.5620}.

The demonstration of these quantum optical phenomena suffers from the low efficiency of generating entangled photon pairs;
because of a low output signal, long-term stability is required for this demonstration. 
To avoid this difficulty, various studies have been proposed to simulate the quantum optical phenomena by using classical light \cite{PhysRevLett.98.223601,kaltenbaek2008quantum,Lavoie2009,PhysRevLett.102.243601,PhysRevA.81.063836,PhysRevA.84.051803,mazurek2013dispersion}.
These studies employ the \textit{time-reversal method}, in which they use the time-reversed version of the conventional schemes using entangled photon pairs.
Because of the time-reversal symmetry of quantum unitary dynamics, a time-reversed experiment reproduces the same result as conventional experiments. 
The time-reversal method replaces the generation of entangled photon pairs with the detection of correlated photon pairs.
Thus we can realize quantum optical phenomena by using only classical light.

The previous investigations employing the time-reversal method realized a time-frequency correlation using chirped-pulse interferometry (CPI) \cite{kaltenbaek2008quantum,Lavoie2009,PhysRevLett.102.243601,PhysRevA.84.051803,mazurek2013dispersion}, which 
 requires a pair of oppositely chirped laser pulses and sum-frequency generation (SFG) to mimic the time-frequency correlation.

In this paper, we observe two-photon phase super-resolution in an unbalanced Michelson interferometer with classical unchirped laser pulses. 
We apply the time-reversal method to a two-photon interference experiment using an unbalanced Michelson interferometer \cite{PhysRevLett.66.1142,PhysRevLett.86.5620}. 
In contrast to CPI, our method does not require chirped laser pulses.
Our scheme extracts the time-frequency correlation by using SFG followed by a bandpass filter.
SFG only converts two photons that arrive simultaneously into a sum-frequency photon, and the bandpass filter extracts sum-frequency photons with a constant frequency.
In the measured interferogram, we observe a classical counterpart of the large difference between the one- and two-photon coherence lengths of entangled photon pairs.
Owing to the simplicity of our experimental setup, we achieved high-visibility two-photon interference with a high degree of efficiency.

This paper is organized as follows.
In Sec.~\ref{sec:theory}, we introduce a theory of conventional and time-reversed methods for observing two-photon phase super-resolution in an unbalanced Michelson interferometer. 
In Sec.~\ref{sec:experiment}, we demonstrate time-reversed experiments for observing two-photon phase super-resolution.
We also experimentally confirm a classical counterpart of the large difference between the one- and two-photon coherence lengths of entangled photon pairs.
In Sec.~\ref{sec:conclusion}, we summarize the findings of our study and discuss the advantages of our method.

\section{Theory}\label{sec:theory}

We first represent the theory of the time-reversal method, which is based on the time-reversal symmetry of quantum unitary dynamics. 
For unitary operator $\hat{U}$ and pair of states $\ket{\mathrm{i}}$ and $\ket{\mathrm{f}}$, relation
\begin{align}
|\bracketii{\mathrm{f}}{\hat{U}}{\mathrm{i}}|^2=|\bracketii{\mathrm{i}}{\hat{U}^{-1}}{\mathrm{f}}|^2\label{eq:5}
\end{align}
holds. 
The relation is easily derived from conjugate transposition $\bracketii{\mathrm{f}}{\hat{U}}{\mathrm{i}}=\bracketii{\mathrm{i}}{\hat{U}^\dag}{\mathrm{f}}^*$ and unitarity $\hat{U}^\dag=\hat{U}^{-1}$.
The left-hand side of Eq.~(\ref{eq:5}) corresponds to the success probability of the projection to final state $\ket{\mathrm{f}}$ for the system evolved with $\hat{U}$ from initial state $\ket{\mathrm{i}}$.
We call this the \textit{time-forward process}.
By utilizing Eq.~(\ref{eq:5}), we can exchange the roles of $\ket{\mathrm{i}}$ and $\ket{\mathrm{f}}$, if we can physically realize time evolution $\hat{U}^{-1}$.
We call the evolution from initial state $\ket{\mathrm{f}}$ by $\hat{U}^{-1}$ projected to final state $\ket{\mathrm{i}}$ the \textit{time-reversed process}.
It has the same success probability as the corresponding time-forward process.

In the following subsections, we illustrate the time-forward and time-reversed processes of the experiment for observing two-photon phase super-resolution in an unbalanced Michelson interferometer.

\subsection{Time-forward process}\label{sec:time-forward-method}

\begin{figure}
\centering
\includegraphics[width=8.5cm]{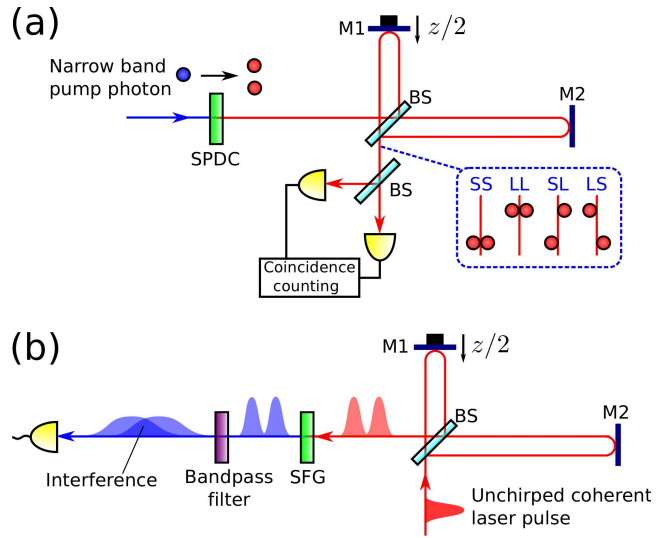}
\caption{(color online). 
Schematic setups for observing two-photon phase super-resolution. 
Michelson interferometer is composed of a 50/50 beam splitter (BS) and mirrors (M1 and M2).
M1 is displaced by $z/2$ to provide optical-path difference $z$ between the two arms of the interferometer. 
(a) Time-forward process.
This setup is composed of a nonlinear optical crystal for SPDC, an unbalanced Michelson interferometer, and two detectors for coincidence counting. 
The dashed-line box describes the possible states of a photon pair.
Letters S and L denote the photons from the shorter and longer arms of the interferometer, respectively. 
When $z$ is large enough, only photon pairs SS and LL contribute to the coincidence counts. 
(b) Time-reversed process. 
This setup is composed of an unbalanced Michelson interferometer, a nonlinear optical crystal for SFG, a bandpass filter, and a detector. 
When $z$ is large enough, two pulses from different arms are individually converted into sum-frequency pulses by SFG. 
After the bandpass filter stretches the pulse widths, these pulses interfere with each other. 
This interference corresponds to that of photon pairs SS and LL in the time-forward process.
}
\label{fig:exp_int_jun}
\end{figure}

The schematic setup of the time-forward experiment is shown in Fig.~\ref{fig:exp_int_jun}(a). 
The input is narrow-band pump photons with center frequency $2\omega_0$. 
The nonlinear optical crystal for SPDC converts the input photons into time-frequency entangled photon pairs with center frequency $\omega_0$. 
After passing through the interferometer, the photon pairs are simultaneously detected by the two detectors.

The measured interference pattern varies depending on optical-path difference $z$ of the interferometer.
Due to the large difference between one- and two-photon coherence lengths $l_1$ and $l_2$ of the entangled photon pairs, one-photon interference occurs for $|z|< l_1$, but two-photon interference occurs for $l_1<|z|<l_2$. In the latter case, we observed two-photon phase super-resolution.

When $|z|<l_1$, the coincidence counting probability per photon pair is given by 
\begin{align}
p(z)=\frac{1}{2}\left[\frac{1+\cos(\omega_0z/c)}{2}\right]^2,\label{eq:3}
\end{align}
where $c$ is the speed of light in a vacuum.
This interference pattern is the square of the one-photon interference pattern.

On the other hand, when $l_1<|z|<l_2$, the two photons from the shorter and longer arms (S and L) are substantially separated in time. 
By setting the time window of the coincidence counting shorter than the arrival-time difference between photons S and L, the photon pairs can be extracted from the same arms (SS and LL) by coincidence counting. 
Since $l_1\ll l_2$ in time-frequency entangled photon pairs, photon pairs SS and LL interfere with each other and generate two-photon interference fringes. 
The coincidence counting probability per photon pair is given by 
\begin{align}
p(z)= \frac{1}{\,8\,}\frac{1+\cos(2\omega_0z/c)}{2},\label{eq:4}
\end{align}
 which indicates two-photon phase super-resolution with perfect visibility. 
The maximum value of Eq.~(\ref{eq:4}) is a quarter of the maximum value of Eq.~(\ref{eq:3}).

\subsection{Time-reversed process}\label{sec:time-reversed-method}

We next present the time-reversal counterpart of the experiment described in the previous subsection. Figure~\ref{fig:exp_int_jun}(b) shows the schematic setup.
The input light is an unchirped coherent laser pulse with center frequency $\omega_0$ and coherence length $l_1'=c\tau_1$, where $\tau_1$ is the pulse duration. 
We describe the complex electric field amplitude of the input light as $E_1(t)=f(t)\ee^{-\ii\omega_0 t}$, where $f(t)$ is the pulse envelope function.
The field amplitude after passing through the interferometer is given by
\begin{align}
E_2(t)&=\frac{1}{2}[E_1(t)+E_1(t+z/c)]\no\\
&= \frac{1}{2}[f(t)+f(t+z/c)\ee^{-\ii\omega_0z/c}]\ee^{-\ii\omega_0 t},
\end{align}
where $z$ is the optical-path difference of the interferometer.
The nonlinear optical crystal for SFG converts the field amplitude into
\begin{align}
E_3(t)&= \alpha E_2(t)^2\no\\
&= \frac{\alpha}{4}[f(t)+f(t+z/c)\ee^{-\ii\omega_0z/c}]^2\ee^{-\ii 2\omega_0 t},
\end{align}
where $\alpha$ is a constant characterizing the SFG efficiency. 
These pulses pass through the bandpass filter followed by a detector.
The bandpass filter narrows the light's bandwidth and broadens its pulse duration.
Assuming $g(t)$ denotes an envelope function of the input light, the effect of a bandpass filter is represented by convolution integral $(g*h)(t)$, where $h(t)$ is the Fourier transform of the transmission spectrum function of the bandpass filter.
Especially when the bandwidth of the transmission spectrum is narrow enough, that is, time width $\tau_2$ of $h(t)$ is much larger than that of input envelope function $g(t)$, envelope function $g(t)$ is approximated as unnormalized delta function $a[g]\delta(t)$, where coefficient $a[g]$ is defined as $a[g]:=\int^\infty_{-\infty}\dd tg(t)$.
Thus the field amplitude after the bandpass filter is described as $(g*h)(t)\approx a[g]h(t)$.
Coherence length $l'_2=c\tau_2$ of the converted pulses is usually much longer than $l'_1$.

Owing to time-reversal symmetry, we expect to observe a similar interferogram as in the time-forward process. 
As calculated below, interference of the pump light occurs when $|z|<l'_1$, but the interference of the sum-frequency light occurs when $l'_1<|z|<l'_2$. 

When $|z|< l_1'$, $f(t)$ and $f(t+z/c)$ almost overlap, $f(t)\approx f(t+z/c)$.
Thus we obtain
\begin{align}
E_3(t)\approx \frac{\alpha}{4}f(t)^2(1+\ee^{-\ii\omega_0z/c})^2\ee^{-\ii 2\omega_0 t}.
\end{align}
The bandpass filter converts $f(t)^2$ into widely spread envelope function $a[f^2]h(t)$.
 The detected signal is described as
\begin{align}
I(z)&\approx \int\dd t\left|\frac{\alpha}{4}a[f^2]h(t)(1+\ee^{-\ii\omega_0z/c})^2\ee^{-\ii 2\omega_0 t}\right|^2\no\\
&=|\alpha|^2\left[\frac{1+\cos(\omega_0z/c)}{2}\right]^2\int\dd t\left|a[f^2]h(t)\right|^2,\label{eq:2}
\end{align}
which reproduces the same interference pattern as the one-photon interference expressed as Eq.~(\ref{eq:3}) in the time-forward process.  
This interference is the square of the white-light interference of the input laser pulses.

On the other hand, when $l_1'<|z|<l_2' $, $f(t)$ and $f(t+z/c)$ only slightly overlap, $f(t)f(t+z/c)\approx0$. Thus we obtain
\begin{align}
E_3(t)\approx \frac{\alpha}{4}[f(t)^2+f(t+z/c)^2\ee^{-\ii2\omega_0z/c}]\ee^{-\ii 2\omega_0 t}.
\end{align}
The bandpass filter converts $f(t)^2$ and $f(t+z/c)^2$ into widely spread envelope functions $a[f^2]h(t)$ and $a[f^2]h(t+z/c)$, respectively.
If $l'_2\gg|z|$, $h(t)$ and $h(t+z/c)$ greatly overlap and are approximated as $h(t)\approx h(t+z/c)$.
The detected signal is described as
\begin{align}
I(z)&\approx \int\dd t\left|\frac{\alpha}{4}a[f^2]h(t)(1+\ee^{-\ii2\omega_0z/c})\ee^{-\ii 2\omega_0 t}\right|^2\no\\
&=\frac{|\alpha|^2}{4}\frac{1+\cos(2\omega_0z/c)}{2}\int\dd t\left|a[f^2]h(t)\right|^2,
\end{align}
which reproduces the same interference pattern as the two-photon interference expressed by Eq.~(\ref{eq:4}) in the time-forward process. 
In this case, the white-light interference of the input pulsed light does not occur due to the large optical path difference of the interferometer. 
After the bandpass filter broadens the pulse widths of the sum-frequency light, these pulses interfere with each other.
We call this \textit{sum-frequency light interference}, which 
exhibits a classical analogue to two-photon phase super-resolution with perfect visibility. 
We can also see that the maximum intensity of the sum-frequency light interference fringes is a quarter of that of the white-light interference fringes.

As seen from the above discussion, the interferogram's shape in the time-reversed process resembles that in the time-forward process. 
The difference between $l'_1$ and $l'_2$ is a classical analogue to the large difference between one- and two-photon coherence lengths $l_1$ and $l_2$.

\section{Experiments and results}\label{sec:experiment}

\begin{figure}
\centering
\includegraphics[width=7cm]{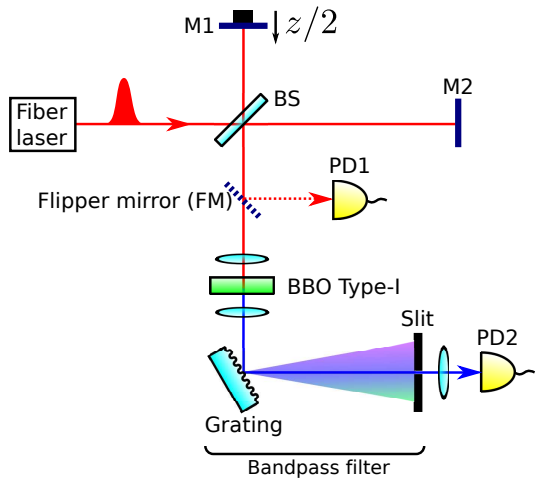}
\caption{(color online). 
Time-reversed experimental setup for observing two-photon phase super-resolution in an unbalanced Michelson interferometer. 
A coherent laser pulse from a femtosecond fiber laser enters the Michelson interferometer. 
To observe the white-light interference fringes, optical-path difference $z$ of the interferometer is adjusted to about zero and the output light is detected by photodiode PD1. 
For observing the sum-frequency light interference, $z$ is adjusted to about 100\,$\upmu$m and the output light is converted into sum-frequency light by a BBO crystal. 
A grating and a slit, both of which function as a bandpass filter, pass the sum-frequency light within a narrow band (0.039\,nm) before the light is detected by a photodiode PD2.}
\label{fig:exp_int_detail}
\end{figure}

We demonstrate a time-reversed experiment for observing two-photon phase super-resolution.
The experimental setup is shown in Fig.~\ref{fig:exp_int_detail}. 
We used a femtosecond fiber laser (Menlo Systems, T-Light 780) with a center wavelength of 782\,nm, a pulse duration of 74.5\,fs FWHM, and an average power of 54.1\,mW. 
The coherence length of the laser pulse was $l_1'=c\tau_1=$ 22.3\,$\upmu$m, where $\tau_1$ is the pulse duration of 74.5\,fs FWHM.
 The Michelson interferometer is composed of a 50/50 nonpolarizing beam splitter for ultrashort pulses (BS) and silver mirrors (M1 and M2). 
 M1 can be translated by several $\upmu$m by a piezoelectric actuator.
The output beam from the interferometer is introduced to the flipper mirror (FM).
To observe the white-light interference, optical-path difference $z$ of the interferometer is adjusted to about zero.  
The output beam of the interferometer is reflected onto FM and detected by a photodiode (PD1; Thorlabs, SM05PD1A). 
On the other hand, for observing sum-frequency light interference, $z$ is adjusted to about 100\,$\upmu$m. 
FM is removed, and the beam is focused by a lens into a 1\,mm-length $\upbeta$-barium borate (BBO) crystal for a collinear type-I SFG.
 The sum-frequency beam is then collimated by another lens and
filtered to pass a 0.039-nm bandwidth centered around 391\,nm by a 3,600\,lines/mm aluminium-coated diffraction grating followed by a slit.
The optical power is measured by a GaP photodiode (PD2; Thorlabs, PDA25K).

\begin{figure}
\centering
\includegraphics[width=6.5cm]{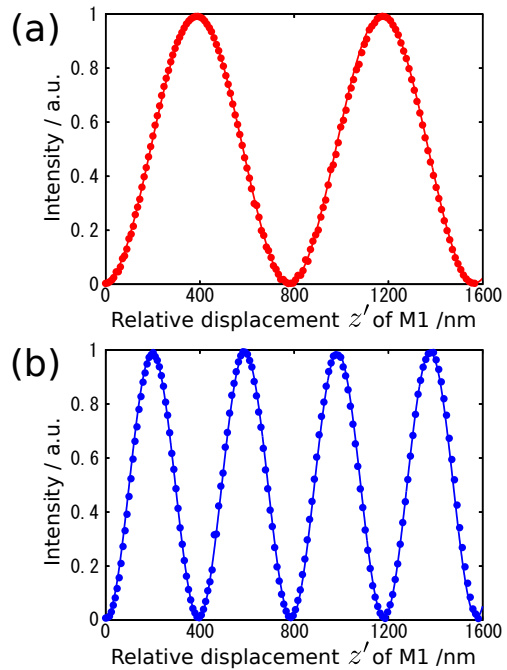}
\caption{(color online). 
Measured interference fringes: (a) white-light and (b) sum-frequency light. 
From the fitted curves in solid lines, visibilities are estimated to be $99.1\%\pm0.2\%$ and $97.9\%\pm0.4\%$, respectively.
Relative displacement $z'$ of M1 depends nonlinearly on piezo voltage $V$.
Assuming that $z'$ is approximated by a quadratic function of $V$, we fitted $z'(V)$ to the measured white-light interference fringes (a).
Calibrated function $z'(V)$ is also applied to the sum-frequency light interference fringes (b). 
Note that $z'$ is the relative displacement of M1 from where optical-path difference $z$ is about zero in (a) and about 100\,$\upmu$m in (b). }
\label{fig:graph_int}
\end{figure}

Figure~\ref{fig:graph_int} shows the measured interference fringes for (a) white-light and (b) sum-frequency light as functions of relative displacement $z'$ of M1. 
The period of the sum-frequency interference is half of the period of the white-light interference. 
This is the classical analogue to the two-photon phase super-resolution. 
The visibilities of the interference fringes were (a) $99.1\%\pm0.2\%$ and (b) $97.9\%\pm0.4\%$, respectively. 
The maximum optical power of the sum-frequency light interference signal was 2.8\,$\upmu$W, which corresponds to about $10^{13}$\,photons/s. 
This count rate is about $10^{11}$ times higher than the two-photon interference signal in the previous time-forward experiments for observing two-photon phase super-resolution in an unbalanced Michelson interferometer \cite{PhysRevLett.66.1142,PhysRevLett.86.5620}.

\begin{figure}
\centering
\includegraphics[width=6.5cm]{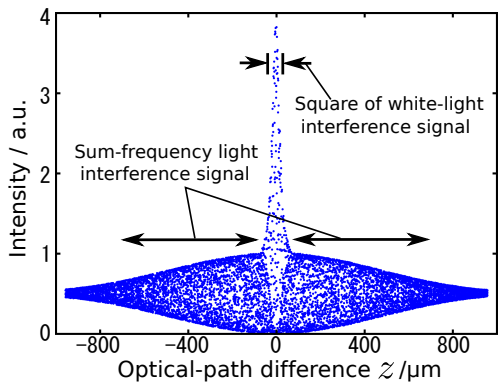}
\caption{(color online). 
Interferogram detected by PD2 for wide range of $z$. 
When $z$ is about zero, the light signal is proportional to the square of the white-light interference signal. 
On the other hand, when $|z|$ is larger than about 100\,$\upmu$m, the light signal exhibits two-photon phase super-resolution. 
The coherence length of the sum-frequency interference is about 22 times larger than that of the white-light interference.
This difference of coherence lengths is a classical analogue to the large difference between one- and two-photon coherence lengths of entangled photon pairs. }
\label{fig:4}
\end{figure}

Next we measured an interferogram detected by PD2 for a wide range of $z$, which is shown in Fig.~\ref{fig:4}.
This experiment confirms the difference between the coherence lengths of the white-light and sum-frequency light interference $l'_1$ and $l'_2$.
In this measurement, M1 is moved by a DC servo motor instead of a piezoelectric actuator in the previous experiment.
We also changed the bandwidth filtered by the grating and the slit to 0.093 nm.
For $z$ near zero, we observed a light signal proportional to the square of the white-light interference signal. 
The measured coherence length of the white-light interference pattern was $23.3\pm0.4$\,$\upmu$m FWHM, which is in good agreement with theoretical coherence length $l_1'=$ 22.3\,$\upmu$m. 
On the other hand, when $|z|$ is larger than about 100\,$\upmu$m, the light signal exhibits two-photon phase super-resolution. 
The measured coherence length of the sum-frequency light interference was $510\pm20$\,$\upmu$m FWHM, which is about 22 times larger than that of the white-light interference. 
The theoretical coherence length is calculated to be $l'_2=(4\ln2/\pi)\lambda^2/\Delta\lambda=1.5$\,mm, where $\lambda=391$\,nm is the central wavelength of the sum-frequency light and $\Delta\lambda=0.093$\,nm is the bandwidth filtered by the grating and the slit.
The measured maximum value of the white-light interference signal is 3.8 times larger than that of the sum-frequency light interference signal. 
The slight shortage compared with the theoretical value of 4 is due to a misalignment of the interferometer. 
This interferogram indicates that our experiment demonstrated a classical analogue to the large difference between the one- and two-photon coherence lengths of entangled photon pairs. 

\section{Summary and Discussion}\label{sec:conclusion}

We observed two-photon phase super-resolution in an unbalanced Michelson interferometer with classical unchirped laser pulses. 
We applied the time-reversal method to a conventional two-photon interference experiment using an unbalanced Michelson interferometer.
The measured interferogram of the experiment exhibits sum-frequency light interference with about 22 times longer coherence length than that of the input laser light.
It is a classical analogue to the large difference between the one- and two-photon coherence lengths of entangled photon pairs.

Kaltenbaek \textit{et al.} \cite{PhysRevLett.102.243601} first observed a classical analogue to the large difference between one- and two-photon coherence lengths using chirped-pulse interferometry.
They used a pulsed light source with an average power of 2.8\,W to observe the white-light and sum-frequency light interferences with visibilities of $87.1\%\pm0.2\%$ and $84.5\%\pm0.5\%$, respectively. 
The maximum optical power of the sum-frequency light interference was about 3.5\,$\upmu$W.
In our experiment the visibilities of the white-light and sum-frequency light interference were $99.1\%\pm0.2\%$ and $97.9\%\pm0.4\%$, respectively.
The maximum optical power of the sum-frequency light interference was 2.8\,$\upmu$W, which is comparable to that of the experiment by Kaltenbaek \textit{et al.}, in spite of our low-power pulsed light source (average power 54.1\,mW).
Such high visibility and high efficiency are due to the simplicity of our experimental setup.

Some previous investigations using coincidence counting for detecting correlated $N$ photons demonstrated $N$-photon phase super-resolution \cite{PhysRevLett.98.223601,PhysRevA.81.063836}.
Coincidence counting, however, cannot detect frequency-correlated photons. 
For this reason, these previous investigations using coincidence counting did not demonstrate the quantum optical phenomena induced by the frequency correlation of photons, such as the difference between one- and two-photon coherence lengths.

We also mention the relation between our experiments and optical lithography. 
Phase super-resolution using entangled photons has been applied to sub-Rayleigh resolution lithography \cite{PhysRevLett.85.2733,PhysRevLett.87.013602}, which is called quantum lithography. 
Sub-Rayleigh resolution lithography was also proposed and demonstrated by a classical optical setup with multiphoton absorption \cite{bentley2004nonlinear,PhysRevLett.96.163603}. 
Pe'er \textit{et al.} \cite{Peer2004} demonstrated sub-Rayleigh resolution lithography using two-photon absorption.
Their experiment resembles ours except it used a Young-like interferometer and detected frequency-correlated photons by two-photon absorption, suggesting 
that some findings about the time-reversal method can be utilized for optical lithography.

%%%%%%

Our study revealed that two-photon phase super-resolution can be realized in a simple classical system without chirped laser pulses.
This simplification enabled us to achieve high-visibility interference with high efficiency.
We expect this technique to open up new practical applications of quantum optical technologies.

\begin{acknowledgments}
This research was supported by the Global COE Program ``Photonics and Electronics Science and Engineering'' at Kyoto University and Grants-in-Aid for Scientific Research No. 22109004 and No. 25287101. One of the authors (S.T.) was supported by JSPS (Grant No. 224850).
\end{acknowledgments}

\appendix

\newpage %Just because of unusual number of tables stacked at end
%\bibliography{ref}% Produces the bibliography via BibTeX.

\end{document}